\documentclass[prl,preprint,aps,showpacs]{revtex4}
\usepackage{epsfig}
\usepackage{graphicx}
\usepackage{epstopdf}

\newcommand{\be}{\begin{equation}}
\newcommand{\ee}{\end{equation}}
\newcommand{\bea}{\begin{eqnarray}}
\newcommand{\eea}{\end{eqnarray}}

\begin{document}
\bibliographystyle{prsty}

\title{A wavepacket approach to periodically driven scattering}
\author{Frank Gro\ss mann}

\affiliation{Institut f\"ur Theoretische Physik, \\
Technische Universit\"at Dresden, \\
D-01062 Dresden, Germany}

\begin{abstract}
For autonomous systems it is well known how to extract 
tunneling probabilities from wavepacket calculations.
Here we present a corresponding approach for periodically 
time-dependent Hamiltonians, valid at all frequencies, field 
strengths, and transition orders. After mapping the periodically driven 
system onto a time-independent one with an additional degree of freedom,
use is made of the correlation function formulation of scattering
[J. Chem. Phys. {\bf 98}, 3884 (1993)]. The formalism is then
applied to study the transmission properties of a resonant
tunneling double barrier structure under the influence of a sinusoidal 
laser field, revealing an unexpected antiresonance in the zero photon 
transition for large field strengths.
\end{abstract}  


\maketitle

\newpage

Recently, the quest for experimental controllability of transport 
processes on the nanoscale has triggered a number of theoretical studies on 
the influence of periodic external fields on molecular scale conductors
\cite{LKHN02,TCD02}
as well as on semiconductor resonant tunneling heterostructures
\cite{Jau90,Cetal90,JW92} and quantum dots \cite{SW96}.
Different theoretical techniques, ranging from Green's
function approaches \cite{Cetal90,JW92,SW96}, to the solution of 
master equations \cite{LKHN02} and wavepacket propagation methods \cite{Jau90}
have been employed to calculate either transmission probabilities
or directly the currents for the considered setups. 

In the seminal work of B\"uttiker and Landauer \cite{BL82}
an analytical approach to calculate the transmission
probability of a sinusoidally driven potential barrier was made.
The perturbative analysis allowed the calculation of
central and side band transmission probabilities at the
incoming energy and at the incoming energy plus or minus the
photon energy. An analogous perturbative way of
extracting those ``driven'' transmission probabilities was shortly 
afterwards used in numerical  \cite{SAL85} and analytical \cite{So88}
studies of resonant  tunneling.
The nonperturbative calculation of tunneling probabilities 
under absorption or emission of photons is a necessity if 
the influence of strong fields on the tunneling process is
to be investigated, however. These probabilities play a central 
role in extending the Landauer formalism to conductance calculations
in the presence of driving
\cite{CLKH03}. A viable generalization of standard approaches
to make the determination of tunneling probabilities for 
arbitrary field strengths possible is therefore needed.

A very intuitive way to calculate the scattering $S(E)$ matrix as
a function of energy,
from which the tunneling probability $T(E)$ can be extracted, has been
given in the framework of reactive scattering theory for autonomous 
systems \cite{TW93}.
This approach relies on the solution of the time-dependent
Schr\"odinger equation and the subsequent Fourier transform
of cross-correlation functions. According to the reactive scattering situation
one distinguishes between internal and translational degrees of freedom,
where the translational one usually represents a tunneling degree
of freedom. In the following, this approach will now be modified to
include, instead of the internal coordinate a new coordinate representing 
time in the case of a periodically time-dependent
system. The absorption or emission of a quantum of radiation can then
be accounted for by calculating the appropriate cross-correlation.
We will elucidate the formalism by applying it to a 
resonant tunneling Hamiltonian modeling e.\ g.\ a weak link metal-metal
contact. 

In order to investigate the phenomenon of tunneling under the influence 
of an external periodic field we follow an S-matrix approach.
Therefore, the Hamiltonian of the driven system with a
periodic driving term of the form $f(x)\sin(\omega t)$ has to be
made explicitly independent of time. This can e.\ g.\ be achieved 
by defining a new pair of canonical variables, $\Theta=\omega t$ and  
$p_\Theta$, 
with the frequency $\omega$ of the external field \cite{LL92}.
The autonomous Hamiltonian for a driven particle of mass $m$
then depends also on the new momentum and coordinate and reads
\be
\label{eq:ham}
H(p_x,x,p_\Theta,\Theta)=\frac{p_x^2}{2m}+V(x)
+f(x)\sin(\Theta)+\omega p_\Theta.
\ee 
This Hamiltonian can easily be quantized by the usual replacements
$p_{x,\Theta}=-i\hbar\partial_{x,\Theta}$. In the following,
we assume that this has been done by keeping in mind the operator
nature of the Hamiltonian 
(in a different notation also known as the ($t,t'$) Hamiltonian \cite{PM93}).

If the coupling term and the potential $V(x)$ are 
localized, i.\ e.\ $V(x),f(x)\to 0$ for $|x|\to \infty$,
then the Hamiltonian at large values of $|x|$ can be written as a sum of 
a kinetic energy term plus a term independent of $x$
\be
H\to H_0(p_x,x,p_\Theta,\Theta)=\frac{p_x^2}{2m}+h(p_\Theta),
\ee 
where the eigenfunctions of $h(p_\Theta)=\omega p_\Theta$ are 
exponentials, in their normalized form given by
\be
\label{eq:chi}
\chi_n(\Theta)=\frac{1}{\sqrt{2\pi}}e^{i n \Theta},
\ee
and living on a finite support $0\leq\Theta < 2\pi$. In the following we
will refer to $n$ as the number of quanta in the external field, due
to the energy eigenvalue $E_n=n\hbar\omega$ corresponding to the 
eigenfunction above. 
The eigenfunctions of the translational part of the Hamiltonian
are energy normalized plane wave states with wavevector $k$.
For the asymptotic Hamiltonian $H_0$ this yields the
direct product eigenstate and energy
\be
\label{eq:en}
\psi_{n,E}=\chi_n(\Theta)
\sqrt{\frac{m}{\hbar k}}\frac{1}{\sqrt{2\pi}}e^{i k x},\qquad
E=n\hbar\omega+\hbar^2 k^2/(2m).
\ee
The eigenstates of the full Hamiltonian $H$ are created 
by applying the M\o ller operator, defined by
$\Omega_{\pm}=\lim_{t\to\mp\infty}\exp\{iHt/\hbar\}\exp\{-iH_0t/\hbar\}$,
to the asymptotic eigenstate
\be
\psi_{n,E}^{\pm}=\Omega_{\pm}\psi_{n,E}
\ee
and serve to define the on-shell S-matrix by
\be
\label{eq:sm}
S_{{\rm R},n';{\rm L},n}(E)\delta(E-E')=
\langle \psi_{{\rm R},n',E'}^-|\psi_{{\rm L},n,E}^+\rangle
\ee
with a slight generalization of notation by including the ``channel'' indices
L,R, denoting to which side of the barrier the eigenstates 
correspond. As noted in \cite{He75}, eigenstates can be
represented by Fourier transform of propagated wavepackets. This idea 
applied to Eq.\ (\ref{eq:sm}) has been used later-on
to extract the scattering matrix using wavepackets \cite{TW93}.
This reasoning can now be taken over to the present situation with the
only difference that the eigenstates corresponding to the internal
motion are states of the field variable $\Theta$ defined in 
Eq.\ (\ref{eq:chi}). The wavefunction to be propagated is a 
Gaussian wavepacket centered in the asymptotic regime of the
translational degree of freedom times an eigenfunction in the field
variable
\bea
\label{eq:wp}
\phi_{{\rm L},n}(x,\Theta)&=&g_{\rm L}(x)\chi_n(\Theta)=
\left(\frac{2\alpha}{\pi}\right)^{1/4}
\nonumber
\\
\exp\{&-&\alpha(x-x_\alpha)^2+ip_\alpha(x-x_\alpha)\}
\chi_n(\Theta).
\eea
For the S-matrix this leads to the expression 
\be
\label{eq:smat}
S_{{\rm R},n';{\rm L},n}(E)=\frac{(2\pi\hbar)^{-1}}
{\eta_{{\rm R},n'}^\ast(E)\eta_{{\rm L},n}(E)}
\int c_{{\rm R},n';{\rm L},n}(t) e^{iEt/\hbar} dt,
\ee 
where a cross correlation function of the form
\be
c_{{\rm R},n';{\rm L},n}(t)=
\langle \phi_{{\rm R},n'}|\exp\{-iHt/\hbar\}|\phi_{{\rm L},n}\rangle,
\ee
is used,
with a final state wavepacket defined analogously to Eq.\ (\ref{eq:wp})
but to the right of the barrier. Furthermore,
the arbitrariness of the initial and final wavefunctions is removed 
from the S-matrix expression by the normalization factor \cite{TW93}
\be
\label{eq:norm}
\eta_{{\rm R},n}(E)=\sqrt{\frac{m}{2\pi\hbar k_n}}\int e^{-ik_n x} 
g_{\rm R}(x) dx,
\ee
with $k_n=\sqrt{2m(E-E_n)}/\hbar$ and a corresponding formula for 
$\eta_{{\rm L},n'}$. 
We stress that one does not need to use M\o ller states
as initial and final wavefunctions as long as they
are located far in the asymptotic regions of the $x$-coordinate,
such that the M\o ller operator becomes the unit operator.
Furthermore, we note in passing, that an approach to extract scattering 
information for periodically driven quantum systems, similar in spirit, 
but based on Floquet theory has e.\ g.\ been pursued in \cite{PM94}, 
\cite{HDR01}.

In the following we calculate transmission probabilities of a
periodically driven system by numerically solving a
2 degree of freedom time-dependent Schr\"odinger equation with Hamiltonian
(\ref{eq:ham}). The $x$ degree of freedom shall correspond to a
tunneling electron coupled to the external field, represented by the
$\Theta$ degree of freedom. Without driving, the electron 
is supposed to move in the symmetric resonant tunneling potential 
$
V(x)=[V_0(1+\exp\{-\beta x_c\})]/[1+\exp\{\beta(|x|-x_c)\}]
-V_{\rm res}\exp\{-\gamma x^2\}
$
introduced by Bringer et al.\ \cite{BHG93} and applied in \cite{BHG93,cp01} 
to study inelastic tunneling in the presence of coupling to
a harmonic oscillator mode. 
We are using the parameter values $V_0=10$ eV, 
$\beta=4$ a.u., $\gamma=1$ a.u., $x_c=4$ a.u. and $V_{\rm res}=14$ eV 
taken from \cite{BHG93} in the following. The undriven potential
is then of the double barrier type, supporting a resonant 
tunneling ``level'' with a 
width of about $0.4$eV at an energy of $E\approx 4.9$eV, well below
the barrier top. 

The periodic field is applied by using the function 
$
f(x)=xV_c\Theta(x_c-|x|)
$ 
localized in $x$ around the double barrier potential. This model is 
motivated by a lead-molecule-lead setup irradiated by a laser field, 
the field-matter interaction being treated in dipole approximation 
and cut off due to the metallic nature of the leads. 
A dipole driving without cutoff could in principle be 
dealt with by applying a Kramers-Henneberger (KH) transformation to 
an oscillatory reference frame \cite{PM94}, \cite{HDR01}. This leads,
however, to a complicated dependence on position {\it and} time 
of the potential $V(x,t)$ and has to be dealt with by a transformation to the
momentum gauge \cite{HDR01}. The field parameters 
are chosen to make the expected effects clearly visible. For the
frequency we thus choose $\hbar\omega=1$ eV, while the strength of
the coupling to the external field $V_c$ will be varied in order to study
its influence on the transmission probabilities.
We will use parameters for which a perturbative approach becomes more
and more questionable.

Numerical results will be presented for the dynamics of 
wavefunction (\ref{eq:wp}) with Gaussian width parameter $\alpha=1$ atomic 
units (a.u.), an initial center of
position $x_\alpha=10$ a.u. and an initial center of momentum
$p_\alpha=0.7$ a.u., having a sufficiently large
overlap with the transmission resonances. We start out by assuming 
that the $\Theta$ degree of freedom
is initially in the $n=0$ state. Although, the number of quanta in the
field has to be very large in order to justify the classical
treatment of the driving term in Eq.\ (\ref{eq:ham}), 
the total quantum wavefunction
can be multiplied with a factor $\exp\{i(m-n)[\Theta-\omega t]\}$
and is still a solution of the time-dependent Schr\"odinger equation
under the Hamiltonian (\ref{eq:ham}). The exponential factor
changes the number of quanta from $n$ to $m$, justifying our 
``shift of the origin'' to $n=0$.
The final states, which the time-evolved wavepacket is to be 
overlapped with, are again direct products as in Eq. (\ref{eq:wp}), but 
located on the right side of the barrier
(with $x_\beta=-x_\alpha$ and $p_\beta=p_\alpha$), with
eigenstates in the $\Theta$ degree of freedom labelled by $n'$.
We stress that there is no restriction on the transition 
(or ``side band'') order $n'$ imposed by our formalism. 
In the following, we focus on the cases $n'=0,1,2$, however.
The grid employed in the numerics has 8192
points in the $x$ direction for the electron and 32 points in the
$\Theta$ direction. In order to avoid unphysical reflections from the
grid boundaries we have employed an absorbing negative imaginary potential
smoothly turned on at large values of $|x|$. 

\begin{figure}
\includegraphics[width=15cm]{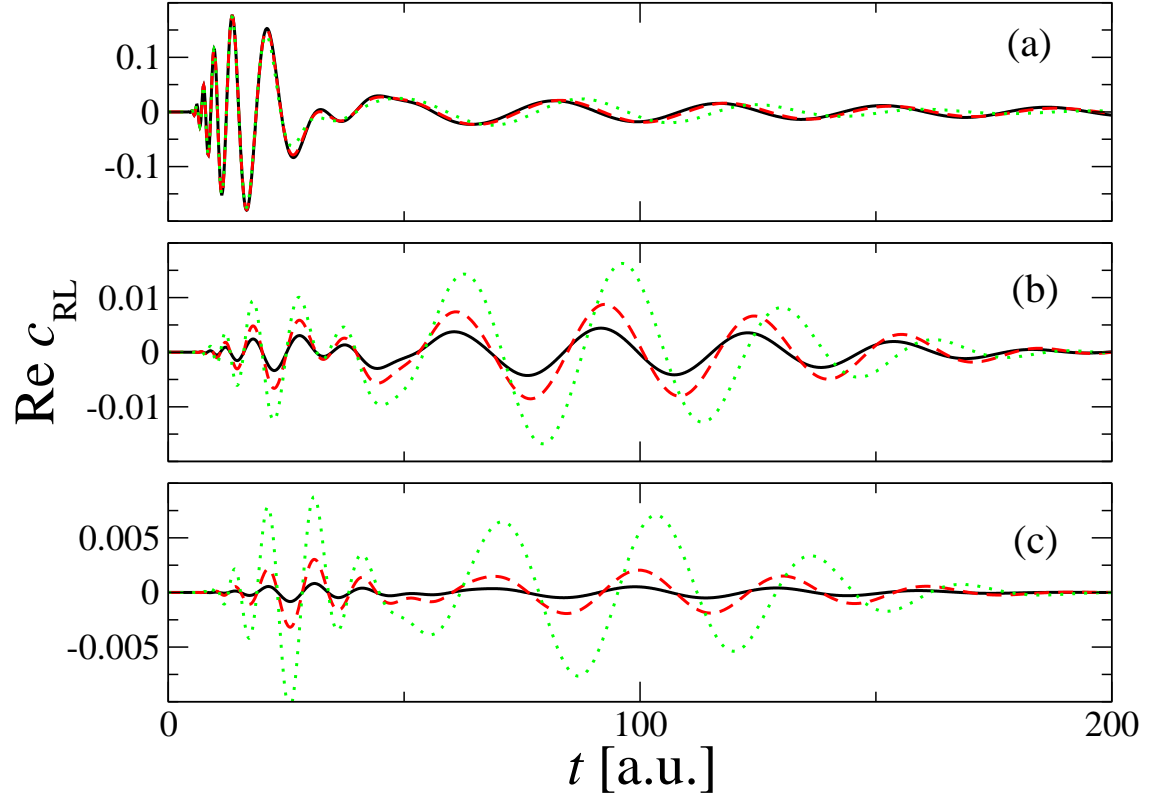}
\caption{\label{fig:cf} 
Real parts of correlation functions $c_{{\rm R},n';{\rm L},n}$ 
for $V_c=0.5$ eV (full line), $V_c=1$ eV (dashed line), 
$V_c=2$ eV (dotted line)
as a function of time in a.u.: (a) $n=n'=0$ (b) $n=0,n'=1$ (c) $n=0,n'=2$}
\end{figure}

In Fig.\ \ref{fig:cf}, the correlation functions entering the 
Fourier transforms in Eq.\ (\ref{eq:smat}) are depicted for 
increasing coupling strengths of
$V_c=0.5,1,2$ eV as a function of time in atomic units.
In panel (a) for the 0-0 transition, it can be seen that the coupling 
to the field leads to a change 
in the long time tail of the oscillation of the correlation function
$c_{{\rm R},0;{\rm L},0}(t)$ setting in at about 50 a.u.. 
For the cases $n'=1,2$, i.\ e.\ excitation of one, respectively two photons 
of the field during the scattering process, the oscillations in the time
signals last longer
than in the $n'=0$ case and die out at times of around 800 a.u. (not shown).
The amplitude of the oscillations is substantially smaller than in panel (a).
The overall shape of the signal in Fig. \ref{fig:cf}(a)
is not changed very much by the coupling.
This does not imply that the resulting transmission will be unaffected,
however. 
In a semiclassical study of undriven 1d-tunneling 
it has e.\ g.\ been shown that, due to the effect of the normalization,
minute differences between the semiclassical and the full quantum
signal can lead to pronounced differences in the corresponding tunneling
probabilities \cite{cpl95}. A similar effect of real, physical origin
can be observed in the following due to the coupling to the light field.

\begin{figure}
\includegraphics[width=15cm]{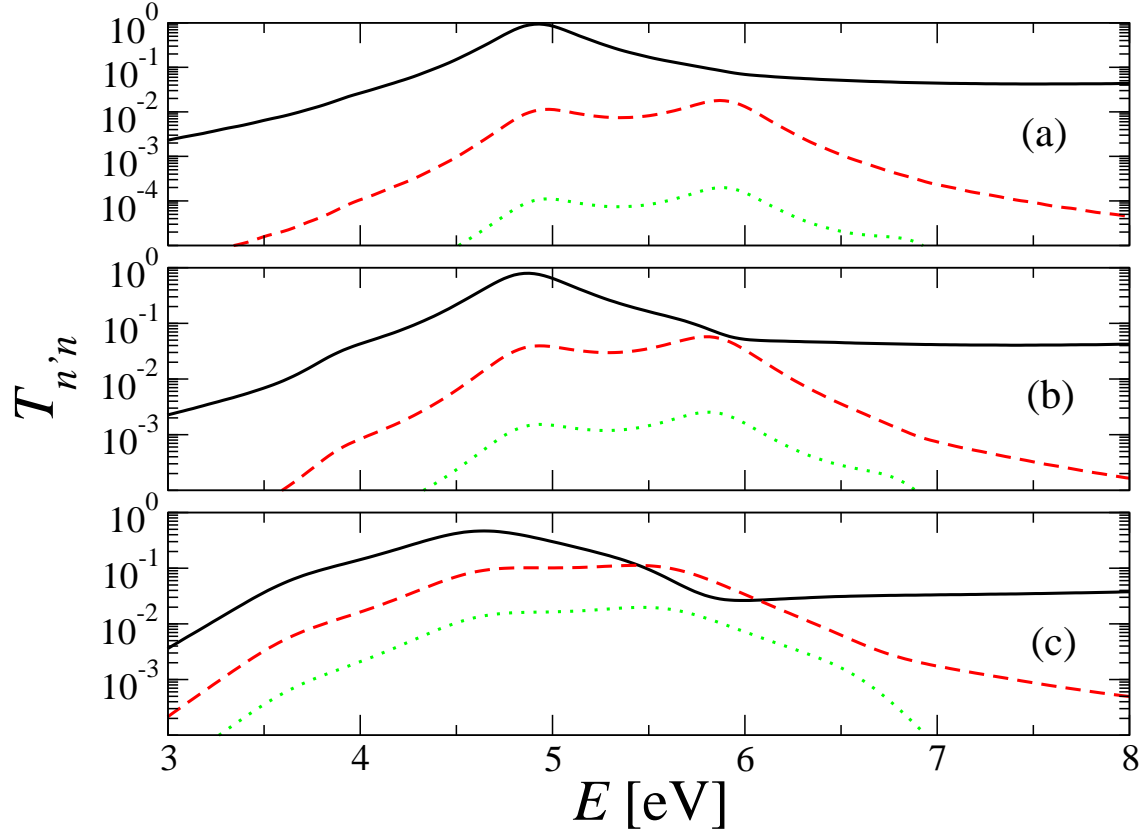}
\caption{Transmission probabilities $T_{n'n}(E)$
for $V_c=0.5$ eV (a), $V_c=1$ eV (b), 
$V_c=2$ eV (c)
as a function of energy (translational equals absolute in the present case)
of the electronic degree of freedom
in eV: $n=n'=0$ (full line), $n=0,n'=1$ (dashed line) $n=0,n'=2$ 
(dotted line).}
\label{fig:t0}
\end{figure}

To this end, we now calculate the tunneling probabilities
from the correlation functions of Fig.\ \ref{fig:cf} according to
\be
\label{eq:tupr}
T_{n'0}(E)=|S_{{\rm R},n';{\rm L},n=0}(E)|^2.
\ee
as a function of (total) energy by using Eq.\ (\ref{eq:smat}) for 
the S-matrix. 
Due to the coupling between the field and
the electron, the final photon number $n'$ may increase (decrease), 
accompanied by
an equivalent energy loss (gain) of the electronic system, which has thus
``emitted'' (``absorbed'') a corresponding number of photons. 
The numerical results are depicted in Fig.\ \ref{fig:t0}. 
For the 0-0 transition, and for $V_c=0.5$ eV, an isolated resonance 
with barely below unit transmission (solid line in panel(a)) at 
around 4.9 eV is observed, close to the unperturbed 
resonance energy.
For increasing coupling strength, the peak in the transmission curve 
acquires a shoulder at about $\hbar\omega$ below the original peak and 
a corresponding dip
to the right (panels (b) and (c)). The transmission probabilities 
with $n'=1,2$ increase with increasing field strength, exhibiting a doubly
peaked structure, with the dominant peak shifted 
from the unperturbed resonance position by $\hbar\omega$
to higher energies and a shoulder blue-shifted by around $2\hbar\omega$
which is most clearly visible for $V_c=$ 1 eV. 
The peak shifted by $\hbar\omega$ describes the scattering
of an electron coming in with an energy higher than the resonance 
but by loosing one quantum of energy closely matching the resonance 
energy leading to an increased transmission probability. 
For $V_c=2$ eV, however, the resonances are already considerably 
broadened having almost lost the  separated
peak structure. Furthermore, the maximum of the
0-0 transition is shifted considerably to the left and the
antiresonance character of the dip at around 6 eV is
clearly emerging. This behaviour can no longer be described and 
understood perturbatively and is in sharp contrast to the 
case of inelastic coupling to a harmonic oscillator \cite{cp01},
where there are no dips but only peaks at integer multiples
of the oscillator frequency to the right of the unperturbed resonance.

Finally, to check consistency, we have studied the case of a single
field quantum being present initially by using $n=1$. As a function
of relative translational energy $E_t=E-n\hbar\omega$
in the incoming channel, the
corresponding transmissions (not shown) then match exactly the
ones shown in Fig. \ref{fig:t0}, i.\ e.\ $T_{11}(E_t)\to T_{00}(E)$
and $T_{21}(E_t)\to T_{10}(E)$. As has been noted before, the field
eigenstate quantum number does not influence
the scattering process. To compare the results directly,
one has to consider the relative
translational energy as the independent variable, however.
Furthermore, in the present case, there is also an equivalence
between e.\ g.\ $T_{01}(E)$ and $T_{10}(E)$ (both as a function of absolute
energy). The special form of the potential together with the driving 
term make the Hamiltonian obey generalized parity symmetry 
(see e.\ g.\ \cite{LKHN03} for a recent discussion of this symmetry 
in a similar context). 
Thus it can be shown by using Eqs.\ (\ref{eq:smat}-\ref{eq:tupr}) that the 
two transmissions have to be equal. We have checked that this is 
indeed the case also in our numerical results ($T_{01}$ not shown). 

We have shown that nonperturbative, exact numerical calculations of 
periodically driven tunneling probabilities, to any transition order, 
can be performed by employing an autonomous Hamiltonian in extended
phase space. Within the wavepacket formalism
of scattering pioneered in \cite{TW93}, the probabilities
for tunneling under absorption or emission of field quanta 
can thus be extracted from suitably chosen cross-correlation functions. 
The presented approach can be applied to all driven scattering
problems, as long as the field coupling $f(x)$ is localized in space. 
A way to deal with pure dipole
driving in a scattering situation has been devised in \cite{HDR01} by 
using a KH transformation which could also be employed in the 
present formalism. 

For resonant tunneling, in sharp contrast 
to the case of inelastic coupling to a harmonic oscillator, 
the field case leads to an antiresonance in the 0-0 transition at high 
frequencies. Furthermore, in addition to the resonant tunneling problem, 
also systems like driven quantum dots \cite{SW96}
or molecular wire ratchets \cite{LKHN02}, and nonresonant barrier 
transmission problems, e.\ g.\ through
semiconductor heterostructures, could be studied
using the new methodology.

The author is indebted to Sigmund Kohler
for valuable discussions
and would like to thank the ``Deutsche Forschungsgemeinschaft''
for financial support through the ``Forschergruppe FOR  335'', project C4.

\newpage


\begin{thebibliography}{10}

\bibitem{LKHN02}
J. Lehmann, S. Kohler, P. H\"anggi, and A. Nitzan, Phys. Rev. Lett. {\bf 88},
  228305  (2002).

\bibitem{TCD02}
A. Tikhonov and R.~D. Coalson, J. Chem. Phys. {\bf 116},  10909  (2002).

\bibitem{Jau90}
A.~P. Jauho, Phys. Rev. B {\bf {\bf 41}},  12327  (1990).

\bibitem{Cetal90}
W. Cai, T.\ F.\ Zheng, P.\ Hu, M.\ Lax, K.\ Shum, and R.\ R.\ Alfano, Phys. Rev. Lett. {\bf {\bf 65}},  104  (1990).

\bibitem{JW92}
P. Johansson and G. Wendin, Phys. Rev. B {\bf {\bf 46}},  1451  (1992).

\bibitem{SW96}
C.~A. Stafford and N.~S. Wingreen, Phys. Rev. Lett. {\bf {\bf 76}},  1916
  (1996).

\bibitem{BL82}
M. B\"uttiker and R. Landauer, Phys. Rev. Lett. {\bf {\bf 49}},  1739  (1982).

\bibitem{SAL85}
A.~D. Stone, M.~Y. Azbel, and P.~A. Lee, Phys. Rev. B {\bf {\bf 31}},  1707
  (1985).

\bibitem{So88}
D. Sokolovski, Phys. Rev. B {\bf {\bf 37}},  4201  (1988).

\bibitem{CLKH03}
S. Camalet, J. Lehmann, S. Kohler, and P. H\"anggi, Phys. Rev. Lett. {\bf 90},
  210602  (2003).

\bibitem{TW93}
D.~J. Tannor and D.~E. Weeks, J. Chem. Phys. {\bf {\bf 98}},  3884  (1993).

\bibitem{LL92}
A.~J. Lichtenberg and M.~A. Lieberman, {\em Regular and Chaotic Dynamics}, 2nd
  ed. (Springer, New York, 1992).

\bibitem{PM93}
U. Peskin and N. Moiseyev, J. Chem. Phys. {\bf {\bf 99}},  4590  (1993).

\bibitem{He75}
E.~J. Heller, J. Chem. Phys. {\bf 62},  1544  (1975).

\bibitem{PM94}
U. Peskin and N. Moiseyev, Phys. Rev. A {\bf {\bf 49}},  3712  (1994).

\bibitem{HDR01}
M. Henseler, T. Dittrich, and K. Richter, Phys. Rev. E {\bf 64},  046218
  (2001).

\bibitem{BHG93}
A. Bringer, J. Harris, and J.~W. Gadzuk, J. Phys.: Condens. Matter {\bf {\bf
  5}},  5141  (1993).

\bibitem{cp01}
F. Grossmann, Chem. Phys. {\bf 268},  347  (2001).

\bibitem{cpl95}
F. Grossmann and E.~J. Heller, Chem. Phys. Lett. {\bf 241},  45  (1995).

\bibitem{LKHN03}
J. Lehmann, S. Kohler, P. H\"anggi, and A. Nitzan, J. Chem. Phys. {\bf 118},
  3283  (2003).

\end{thebibliography}

\newpage

\newpage

\end{document}